\documentclass[sigconf,natbib=false]{acmart}

\usepackage[utf8]{inputenc}
\usepackage{graphicx}
\usepackage{xcolor}
\usepackage{svg}
\usepackage{booktabs}
\usepackage{multirow}
\usepackage{amsmath}
\usepackage{algpseudocode}
\usepackage{listings}
\usepackage{subcaption}
\usepackage{comment}

\usepackage[binary-units=true]{siunitx}

\usepackage[absolute,overlay]{textpos}

\makeatletter
\def\@ACM@checkaffil{
    \if@ACM@instpresent\else
    \ClassWarningNoLine{\@classname}{No institution present for an affiliation}%
    \fi
    \if@ACM@citypresent\else
    \ClassWarningNoLine{\@classname}{No city present for an affiliation}%
    \fi
    \if@ACM@countrypresent\else
        \ClassWarningNoLine{\@classname}{No country present for an affiliation}%
    \fi
}
\makeatother

\setcopyright{none}
\renewcommand\footnotetextcopyrightpermission[1]{}

\usepackage[style=acmnumeric,citestyle=numeric-comp,backend=biber,sorting=none]{biblatex}
\begin{document}

\title{POSTER: spaceQUIC: Securing Communication in Computationally Constrained Spacecraft}

\author{Joshua Smailes}
\authornotemark[2]
\email{joshua.smailes@cs.ox.ac.uk}
\affiliation{%
    \institution{University of Oxford}
}

\author{Razvan David}
\authornotemark[2]
\email{razvan.david@cs.ox.ac.uk}
\affiliation{%
    \institution{University of Oxford}
}

\author{Sebastian K{\"o}hler}
\email{sebastian.kohler@cs.ox.ac.uk}
\affiliation{%
    \institution{University of Oxford}
}

\author{Simon Birnbach}
\email{simon.birnbach@cs.ox.ac.uk}
\affiliation{%
    \institution{University of Oxford}
}

\author{Ivan Martinovic}
\email{ivan.martinovic@cs.ox.ac.uk}
\affiliation{%
    \institution{University of Oxford\\\,}
}

\begin{abstract}
Recent years have seen a rapid increase in the number of CubeSats and other small satellites in orbit -- these have highly constrained computational and communication resources, but still require robust secure communication to operate effectively.
The QUIC transport layer protocol is designed to provide efficient communication with cryptography guarantees built-in, with a particular focus on networks with high latency and packet loss.
In this work we provide \textit{spaceQUIC}, a proof of concept implementation of QUIC for NASA's ``core Flight System'' satellite operating system, and assess its performance.
\end{abstract}

\maketitle

\begin{textblock*}{7cm}(1.8cm,7.8cm)
  \small{\textdagger\;Both authors contributed equally to this work.}
\end{textblock*}

\vspace{-1.5em}

\section{Motivation}\label{sec:motivation}

Alongside a general upward trend in the number of satellites in orbit, there has been a marked recent increase in the number of small satellites in space.
This increase has been driven by a number of factors, including a growing availability of cheap Commercial Off-The-Shelf (COTS) components, satellite ride-sharing (in which smaller satellites are launched alongside a larger payload), and the rise in popularity of the CubeSat, a standardized design enabling cheap ride-sharing.

Another significant factor has been software availability -- access to open-source operating systems and libraries allow operators to focus on building payload-specific functionality, reducing unnecessary mission development.
The \textit{core Flight System} (cFS) is a popular open-source satellite operating system built by NASA from historical missions, and is used in many ongoing and planned missions~\cite{mccomasCore2015}.
It is actively maintained by NASA and the open-source community surrounding the project, and is easily extensible through the addition of libraries or apps to support specific payloads or ancillary functions.
It is built on top of an Operating System Abstraction Layer (OSAL), making it easy to port to new hardware platforms, alongside those for which it is already supported.
For these reasons it is a popular choice of satellite operating system and is used in a wide range of CubeSat missions.

One key challenge in space systems is performant secure communication -- the vast majority of communication occurs over radio signals which are subject to significant path loss, atmospheric noise, and multipath distortion.
This problem is exacerbated in CubeSats, which often use omnidirectional antennas due to an inability to orient themselves, or antennas with lower gain due to their limited size.
As a result, much of these satellites' communication is low throughput and subject to data loss or corruption.

The TCP transport layer protocol is rarely used -- its congestion control algorithm assumes that packet loss is caused by a congested link and waits for retransmission.
These assumptions do not apply to point-to-point satellite communications, and result in a significant decrease in throughput.
Instead, datagram-oriented protocols like UDP or the Space Packet Protocol (SPP) are used.
These are secured using symmetric cryptography through the Space Data Link Security (SDLS) protocol, implemented in the \textit{CryptoLib} cFS library~\cite{ccsdsSpace2015}.
This provides fewer security guarantees than asymmetric cryptography, but requires less computational overhead.

The QUIC transport layer protocol, introduced in 2012, addresses these concerns~\cite{iyengarQUIC2021}.
The connection establishment process is highly streamlined, and in many cases data can start being sent with 0 round trips of setup.
The protocol also provides improved congestion control and recovery from losses, resulting in significantly better throughput over lossy and noisy connections.
Furthermore, security is built into the protocol, providing all the security guarantees of an asymmetric cryptosystem with at most one network round-trip time of setup.
These factors make QUIC highly attractive in the context of lossy satellite connections -- existing work has shown that combining QUIC with performance enhancing proxies can provide better performance and security over satellite internet connections~\cite{pavurQPEP2021}.
However, there is currently no way to leverage the benefits of QUIC when in direct communication with satellites.

\vspace{-1em}
\subsection{Contributions}\label{sec:contributions}

In this work we introduce the \textit{spaceQUIC} library, implementing QUIC functionality on the cFS satellite operating system.
This brings the additional performance and resilience of QUIC to space missions, alongside the increased security of asymmetric cryptography.
This library can be used as a replacement for \textit{CryptoLib}, the existing library providing asymmetric cryptography through the SDLS protocol.

We provide a high-level overview of the cFS architecture and describe how \textit{spaceQUIC} fits into this model.
We also explain how to extend \textit{spaceQUIC} to work with real-world missions.

All code has been made open source under the Apache 2.0 license, and can be found at \url{https://github.com/ssloxford/spaceQUIC}.
For ease of setup, we also provide an instance of cFS preconfigured to use \textit{spaceQUIC}, both as source code and a Docker container.

\section{Architecture}\label{sec:architecture}

\begin{figure}
\centering
\includegraphics[width=\columnwidth]{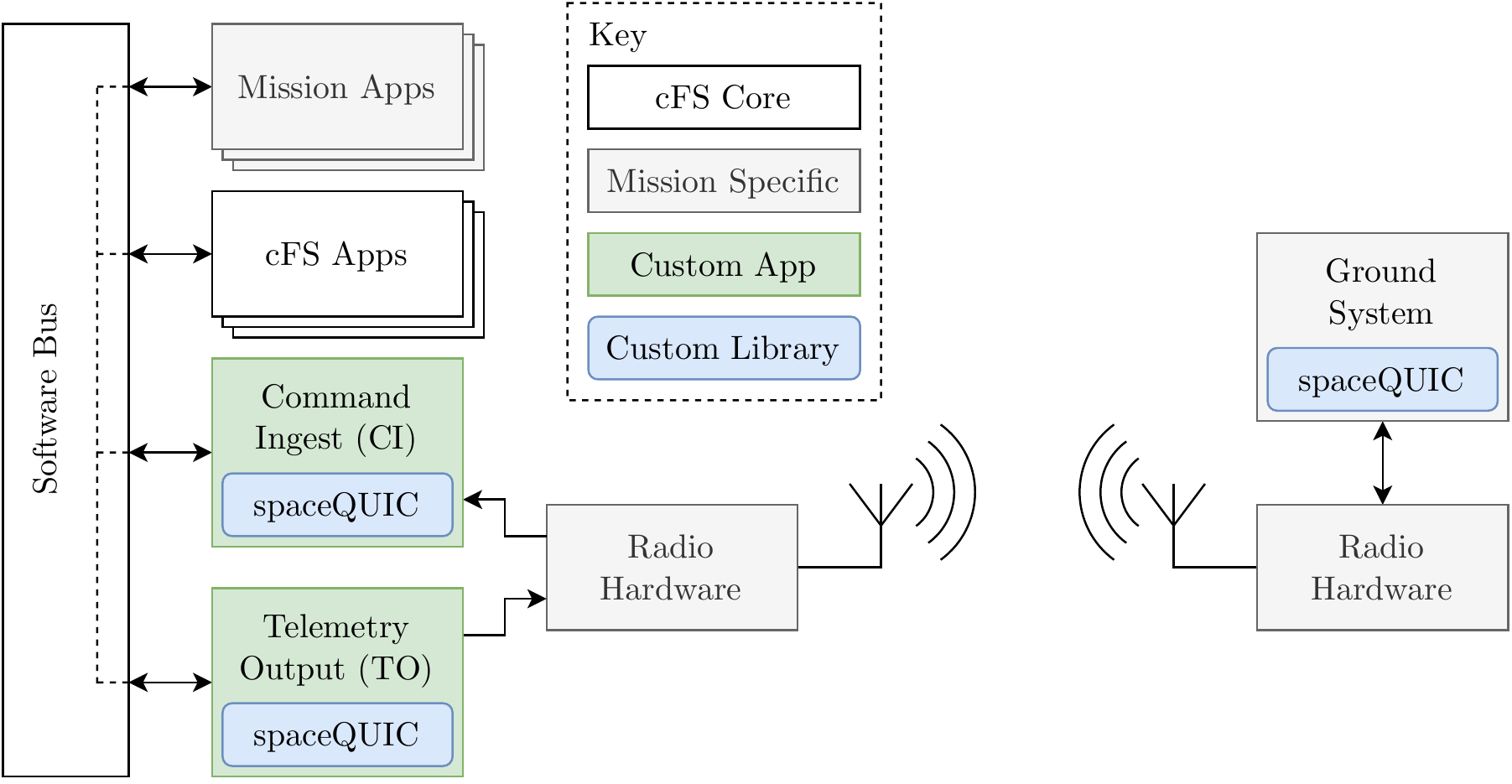}
\caption{The overall architecture of a space system running cFS using \textit{spaceQUIC}.}
\label{fig:cfs-communication}
\vspace{-1em}
\end{figure}

Figure~\ref{fig:cfs-communication} shows the structure of a cFS system using \textit{spaceQUIC}.
Thanks to the modular structure of cFS, both central cFS functionality, as well as mission-specific applications and libraries, are left unchanged.
All communication occurs over a central software bus, with applications exchanging data through a publish/subscribe message passing system.
This means the underlying communication stack is abstracted away from most of the system, with data sent only via the software bus.

\textit{spaceQUIC} is provided as a cFS library, giving access to the required QUIC functionality.
This library is used by modified Command Ingest (CI) and Telemetry Output (TO) applications -- these are provided as part of the standard cFS system, and are used for processing commands, and sending telemetry and housekeeping data back to the ground system.
The provided ``lab'' versions of these applications send data directly over the network, and are modified on a per-mission basis to support that mission's radio hardware.
Existing CI and TO applications can also be modified to support \textit{spaceQUIC}, replacing calls to \textit{CryptoLib} or other security/networking libraries.

The \textit{spaceQUIC} library supports two implementations of \mbox{SSL}/\mbox{TLS}: OpenSSL and WolfSSL~\cite{wolfssl}.
WolfSSL is designed for use in embedded devices and optimized to minimize resource usage, making it ideal for small satellites.

\section{Performance}\label{sec:performance}

\begin{table}
    \centering
    \sisetup{table-number-alignment=right}
    \caption{Overall memory usage of cFS under each security configuration.}
    \label{tab:memory-usage}
    \resizebox{\columnwidth}{!}{
        \begin{tabular}{lS[table-format=3.1]S[table-format=2.1]}
            \toprule
            Security       & {Peak heap usage (\SI{}{\kilo\byte})} & {Peak RSS usage (\SI{}{\mega\byte})} \\
            \midrule
            None           & 84.5                                  & 6.7                                  \\
            SDLS           & 89.5                                  & 8.3                                  \\
            QUIC (OpenSSL) & 583.3                                 & 13.3                                 \\
            QUIC (WolfSSL) & 344.8                                 & 9.5                                  \\
            \bottomrule
        \end{tabular}
    }
\end{table}

\begin{figure}
    \centering
    \includegraphics[width=\columnwidth]{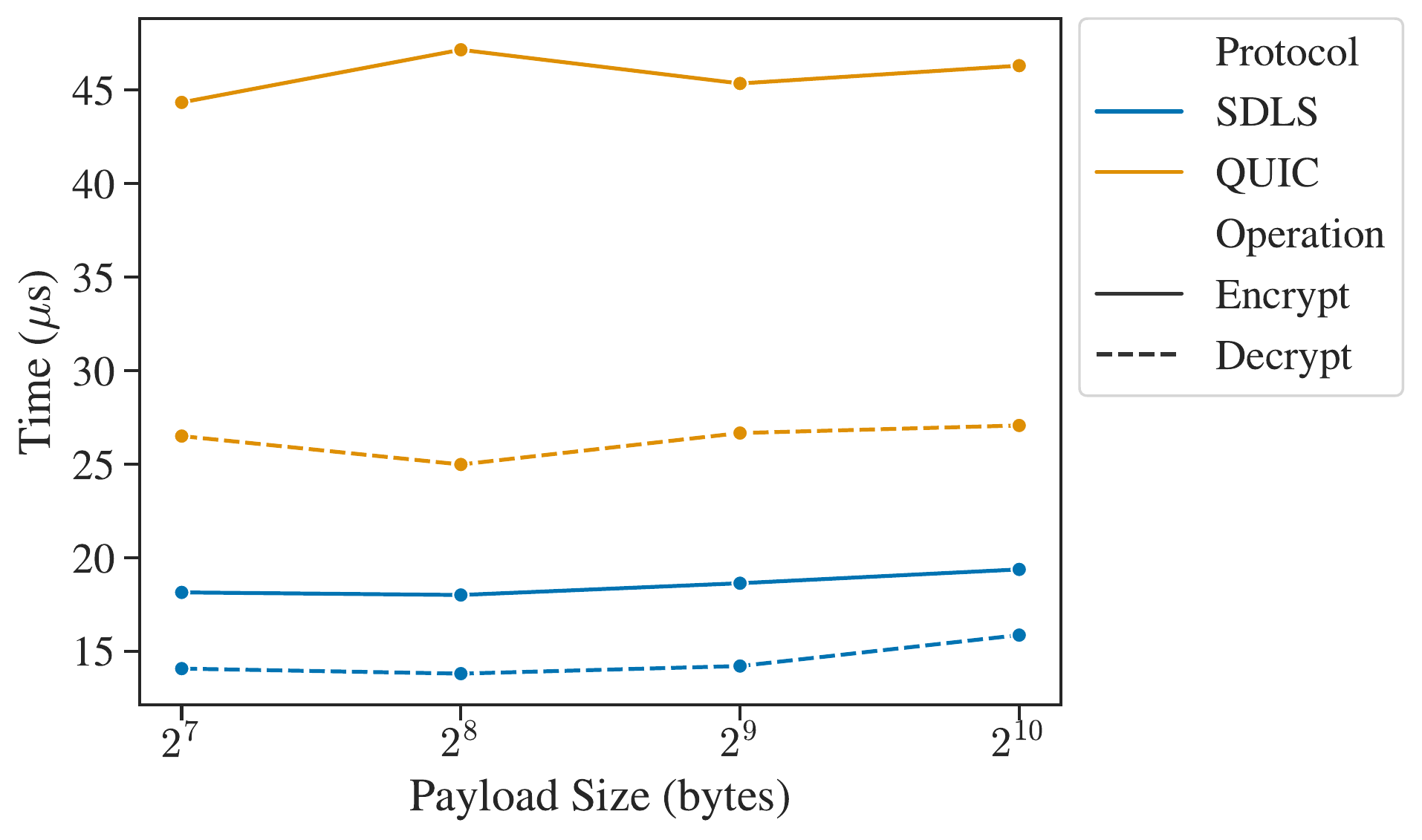}
    \caption{Encryption and decryption times for SDLS and QUIC.}
    \label{fig:enc_dec}
\end{figure}

In this section we assess the performance and resource usage of \textit{spaceQUIC} to demonstrate its usefulness in embedded contexts.

All experiments were performed on a Dell laptop with an Intel i7-8750H CPU and 16GB of DDR4 memory, limited to a single thread.
Embedded hardware comparable to onboard satellite hardware was not available, so we focus on relative performance.

Table~\ref{tab:memory-usage} shows the memory usage of cFS under each configuration, looking at both peak heap usage and Resident Set Size (RSS) of the process.
We see that QUIC uses significantly more heap space than SDLS, but when using WolfSSL overall memory usage is only slightly increased.
It is likely that there would be no memory usage problems running QUIC on all but the most computationally constrained spacecraft.

We also measured execution time, seen in Figure~\ref{fig:enc_dec}.
From these results we observe that QUIC is $1.5$ to $2.5$ times slower than SDLS -- this is unsurprising due to the greater requirements of asymmetric cryptography, but further testing on embedded hardware is needed.

Further testing is also also required to measure performance when latency and packet loss are high -- due to the protocol's design, \textit{spaceQUIC} is likely to perform well in these scenarios.

\printbibliography

\end{document}